\documentclass[aps,prl,twocolumn,groupedaddress]{revtex4}

\usepackage{amsmath}
\usepackage{graphicx}
\usepackage{color}
\usepackage{epstopdf}

\usepackage{ulem}

\begin{document}

\title{Super-Diffusive Spin-Transport as a Mechanism of Ultrafast Demagnetization}

\author{M. Battiato}
\email[]{marco.battiato@fysik.uu.se}
\author{K. Carva}
\author{P.\ M. Oppeneer}
\affiliation{Dept.\ of Physics and Astronomy, Uppsala University, Box 516, SE-75120 Uppsala, Sweden}

\date{\today}

\begin{abstract}
 We propose a semi-classical model for femtosecond-laser induced demagnetization due to  spin-polarized excited electron diffusion in the super-diffusive regime. Our approach treats the finite elapsed time and transport in space between multiple electronic collisions exactly, as well as the presence of several metal films in the sample. Solving the derived transport equation numerically we show that this mechanism accounts for the experimentally observed demagnetization within 200\,fs in Ni, without the need to invoke any angular momentum dissipation channel.
\end{abstract}

\pacs{?}

\maketitle

 Excitation with femtosecond laser pulses is known for more than a decade to cause an ultrafast quenching of the magnetization in metallic ferromagnets \cite{beaurepaire96}. The achieved demagnetization times are typically 100-300 fs for ferromagnets such as Ni \cite{beaurepaire96,stamm07}. Hence, laser-induced demagnetization opens up new, interesting routes for magnetic recording with hitherto unprecedented speeds \cite{vahaplar09}. However, in spite of the technological importance the mechanism underlying the femtosecond demagnetization remains highly controversial. A common belief is that there should exist an ultrafast channel for the dissipation of spin angular momentum \cite{koopmans05,koopmans09,zhang09,bigot09,krauss09}. Several such mechanisms through which an excited electron can undergo a spin-flip in a ferromagnetic metal are currently being debated. The main proposed mechanisms for a fast spin-flip process are a Stoner excitation, an inelastic magnon scattering, an Elliott-Yafet-type of phonon scattering \cite{koopmans05,koopmans09},  spin-flip Coulomb scattering \cite{krauss09}, laser-induced spin-flips  \cite{zhang09}, or relativistic quantum electrodynamic processes \cite{bigot09}. An effect that, until recently \cite{malinowski08}, has been regarded to play only a marginal role is the spin-polarized transport of laser-excited hot electrons. 
 	
 In this Letter we show that spin-dependent transport of laser-excited electrons provides a considerable contribution to the ultrafast demagnetization process and can even completely explain it. We demonstrate this by developing a transport equation for the super-diffusive flow of spin-polarized electrons. A few approaches to describe the electron motion have been attempted previously \cite{knorren00,ludeke93}. In our theory, however, we take into account the {\it whole} process of multiple, spin-conserving electron scattering events and electron cascades created by inelastic electron scattering. Also the presence of different metallic films in the probed material is treated. We solve the developed theory numerically for ferromagnetic Ni, for which the femtosecond demagnetization is well documented \cite{beaurepaire96,cheskis05,stamm07}, and show that  a large demagnetization in a few hundred femtoseconds is generated.
	
\begin{figure}
 \includegraphics[width=0.4\textwidth]{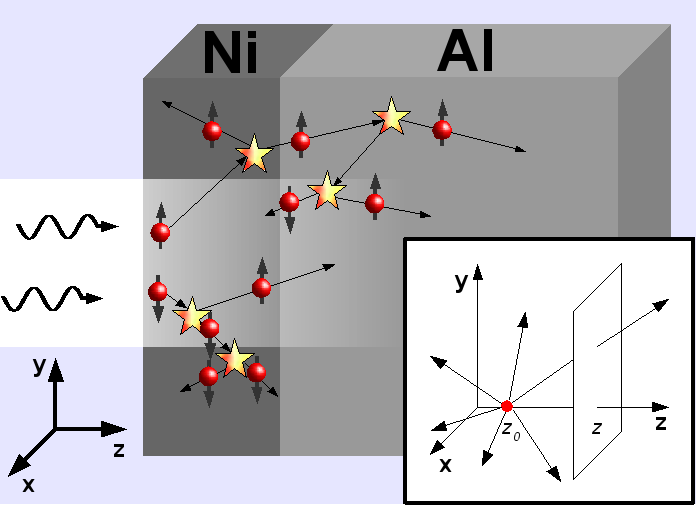}
 \caption{ (Color online) Sketch of the super-diffusive processes caused by  laser excitation. Different 
 mean-free-paths for majority and minority spin carriers are shown and also the generation of a cascade of electrons after an inelastic scattering (elastic scatterings are not shown for simplicity). The inset shows the geometry for the calculation of the electron flux term in the continuity equation.}\label{fig:Ball_diff}
\end{figure}


 The typical geometry for a femtosecond laser experiment is depicted in Fig.\ \ref{fig:Ball_diff}. The intense laser beam creates excited hot electrons in the ferromagnetic film, which will start to move in a random direction. Our goal is to compute the time-dependent magnetization resulting from the super-diffusive motion in the laser spot. Due to the fact that the electronic mean-free-path (up to a few tens of nm) is much smaller than the diameter of the laser spot normally used, see e.g. \cite{stamm07}, we developed a uniaxial model where only the $z$ dependence is kept, but the formalism can easily be applied to describe systems with less symmetry.

 After the absorption of a photon, an electron will be excited typically from a $d$-band to the $sp$-like bands above the Fermi level. The mobility of $sp$-like electrons is much larger than that of $d$ electrons. Therefore in the following we treat the $d$ electrons as quasi-localized and compute the transport only for the mobile $sp$-like electrons. We furthermore use that the optical excitation is spin conserving.  Because of the very small linear momentum carried by a photon the angular probability density of the emission direction can be safely considered isotropic over all solid angles.  Hence, the excited electron will start moving in a random direction. The outgoing trajectory is treated as a straight line up to the first scattering event. As the derivation involves many variables, in the following we show explicitly only those that are relevant at that moment.
 We start with describing the motion of electrons before the first scattering event, which (for simplicity) we call first-generation electrons. The trajectory $s(t)$ for a first-generation electron is given implicitly by $ \int_0^s ds'/ v \left( z\left(s'\right) \right) = t$, where $v\left(\sigma,E,z\right)$ is the velocity, $z\left(s\right)$ the $z$-coordinate of the particle when the coordinate on the trajectory is $s$, and $t$ the time needed to reach $s$. The probability to reach a point $s$ without being scattered is $P\left(s\right)=exp\left[ - \int_0^s ds' / \tau\left(z\left(s'\right)\right)v\left(z\left(s'\right)\right) \right]$, where $\tau\left(\sigma,E,z\right)$ is the lifetime.  The velocity and lifetime are both considered dependent on the position $z$ (therefore on the material) and on the spin $\sigma$ and energy $E$ but not on the direction of motion.

 Using that the probability of the emission direction is isotropic and integrating over all possible emission angles, we compute the statistically averaged flux $\phi$ at a time $t$ through a surface located at $z$ due to an electron that starts its movement at $z_0$ at time $t_0$ (see inset in Fig.~\ref{fig:Ball_diff}),
	\begin{eqnarray}
	\label{eq:flux_3D}
	 \phi\left(z,t;z_0,t_0\right) &=  \frac{\widetilde{\left[\Delta t\right]}}{2\left(t-t_0\right)^2} 
	 \left(exp\left[- \widetilde{\left[\frac{\Delta t}{\tau}\right]}/\widetilde{\left[\Delta t\right]} \right]\right)^{\left(t-t_0\right)}   \times \nonumber \\
&	  \Theta \left[\left(t-t_0\right)-|\widetilde{\left[\Delta t\right]} | \right] ,  
	\end{eqnarray}
 where we defined
	\begin{equation}
		\widetilde{\left[\frac{\Delta t}{\tau}\right]}=
		\int_{z_0}^{z} \frac{dz'}{\tau\left(z'\right)v\left(z'\right)}, ~~~~
	\widetilde{\left[\Delta t\right]}=
		\int_{z_0}^{z} \frac{dz'}{v\left(z'\right)} ,
	\end{equation}
 and used that an electron which suffers a scattering does not give any contribution to the first-generation flux.
	
 If, instead of exciting a single electron, a distributed source of excited electrons is present, the total first-generation flux due to all electrons with spin $\sigma$ and energy $E$ is
	\begin{equation}
	 \Phi  \left(z,t\right)
	  = \int_{-\infty}^{+\infty}\!\!\! dz_0 \int_{-\infty}^{t}\!\!\! dt_0 \: S^{ext} \left(z_0,t_0\right) \phi\left(z,t;z_0,t_0\right) ,
	\end{equation}
 where $S^{ext}=S^{ext}\left(\sigma,E,z,t\right)$ is the electron source term that has to be computed from the spatial and temporal profile of the laser and the absorption probability. We define the operator $\hat{\phi}$,  by $\hat{\phi} S^{ext} \equiv \Phi $.
	
 Once we have the expression for the flux it is straightforward to write the continuity equation for the density of first-generation electrons of given spin and energy, $n^{\left[1\right]}\left(\sigma,E,z,t\right)$,
	\begin{equation}
	\label{eq:continuity_equation_1_scattering}
	 \frac{\partial n^{\left[1\right]}}{\partial t}+\frac{n^{\left[1\right]}}{\tau} = -\frac{\partial \hat{\phi} S^{ext}}{\partial z}+ S^{ext}.
	\end{equation}
 The second term is the scattering term that acts as a reaction term removing electrons from the density when they are scattered. 

 To describe the motion of electrons after the first scattering event, we make the assumption that the angular probability density of the emission is again isotropic and uncorrelated to the incoming direction. In the case of scattering with phonons, impurities, and other large mass particles this is almost exactly true. Also for scattering of $sp$-like electrons with the much heavier $d$-electrons it is a suitable approximation since it leads to an underestimation of the diffusive process. The density of second-generation electrons, $n^{\left[2\right]}\left(\sigma,E,z,t\right)$, is described by Eq.\ (\ref{eq:continuity_equation_1_scattering}), too,  when instead of using $S^{ext}$ we use $S^{\left[2\right]}\left(\sigma,E,z,t\right)$, defined as:
	\begin{equation}
	S^{\left[2\right]}=\sum_{\sigma'}\int  p\left(\sigma,\sigma',E,E',z\right) \frac{n^{\left[1\right]}\left(\sigma',E',z,t\right)}{\tau\left( \sigma',E',z \right)} dE' ,
	\end{equation}
 which is the scattering term coming from the first-generation weighted by the transition probability after a scattering, $p\left(\sigma,\sigma',E,E',z\right)$. The latter includes the effect of both elastic and inelastic scatterings as well as the generation of cascade electrons. We note that in principle the transition probability could handle spin-flip events but we assume them to be negligible. We define the operator $\hat{S}$ as $\hat{S}  n^{\left[1\right]} \equiv S^{\left[2\right]}$.  Applying the same procedure we  obtain the equation for the density of the third-generation electrons $n^{\left[3\right]}$, and so on. Summing up everything we derive a set of coupled transport equations,
	\begin{equation} \label{eq:continuity_eq_total_density}
	 \frac{\partial n^{tot}}{\partial t}+\frac{n^{tot}}{\tau}= \left(-\frac{\partial }{\partial z} \hat{\phi} + \hat{I}\right) \left(\hat{S}n^{tot}+S^{ext}\right).
	\end{equation}
 $\hat{I}$ is the identity operator.
	
 Eq.\ (\ref{eq:continuity_eq_total_density}) describes the fast transport of laser-excited electrons. A major question is whether it can explain the femtosecond demagnetization observed in pump-probe magneto-optical experiments. Before addressing this question we stress that the developed transport process is both different from ballistic and diffusive transport. Standard diffusive processes that are governed by Brownian motion are characterized by the variance of the displacement of a particle distribution $\sigma^2$ which grows linearly with time: $\sigma^2\left(t\right)=t^{\gamma}$, with $\gamma$=1  \cite{Metzler00}. Ballistic diffusion is characterized by $\gamma$=2. The here-developed electron motion description is in the category of super-diffusive processes ($\gamma$$>$1) with the further distinction that $\gamma$ is time-dependent and goes from a ballistic regime $\gamma$=2 for small times to normal diffusion $\gamma$=1 for long times. This furthermore emphasizes that a standard diffusion model is inapplicable to electron motion on the fs timescale.

 Super-diffusive transport may give rise to demagnetization, because, first, laser-excited electrons in $sp$-bands have high velocities  (about 1~nm/fs), and second, excited spin majority and minority electrons have different lifetimes. The latter lends excited majority carriers in typical $3d$ ferromagnets a high mean-free-path whereas minority carriers are much less mobile. This may lead to a depletion of majority carriers in the magnetic film and a transfer of magnetization away from the surface. Moreover, an excited electron experiencing an inelastic scattering with another electron will transfer part of its energy to the other one, generating an electron cascade. The newly excited electron will, now, have enough energy to contribute with its motion to the demagnetization. The occurring energy transfer is computed here from the classical treatment of a two-particle collision.

\begin{figure}
 \includegraphics[trim = 3mm 0mm 20mm 17mm, clip, width=0.45\textwidth]{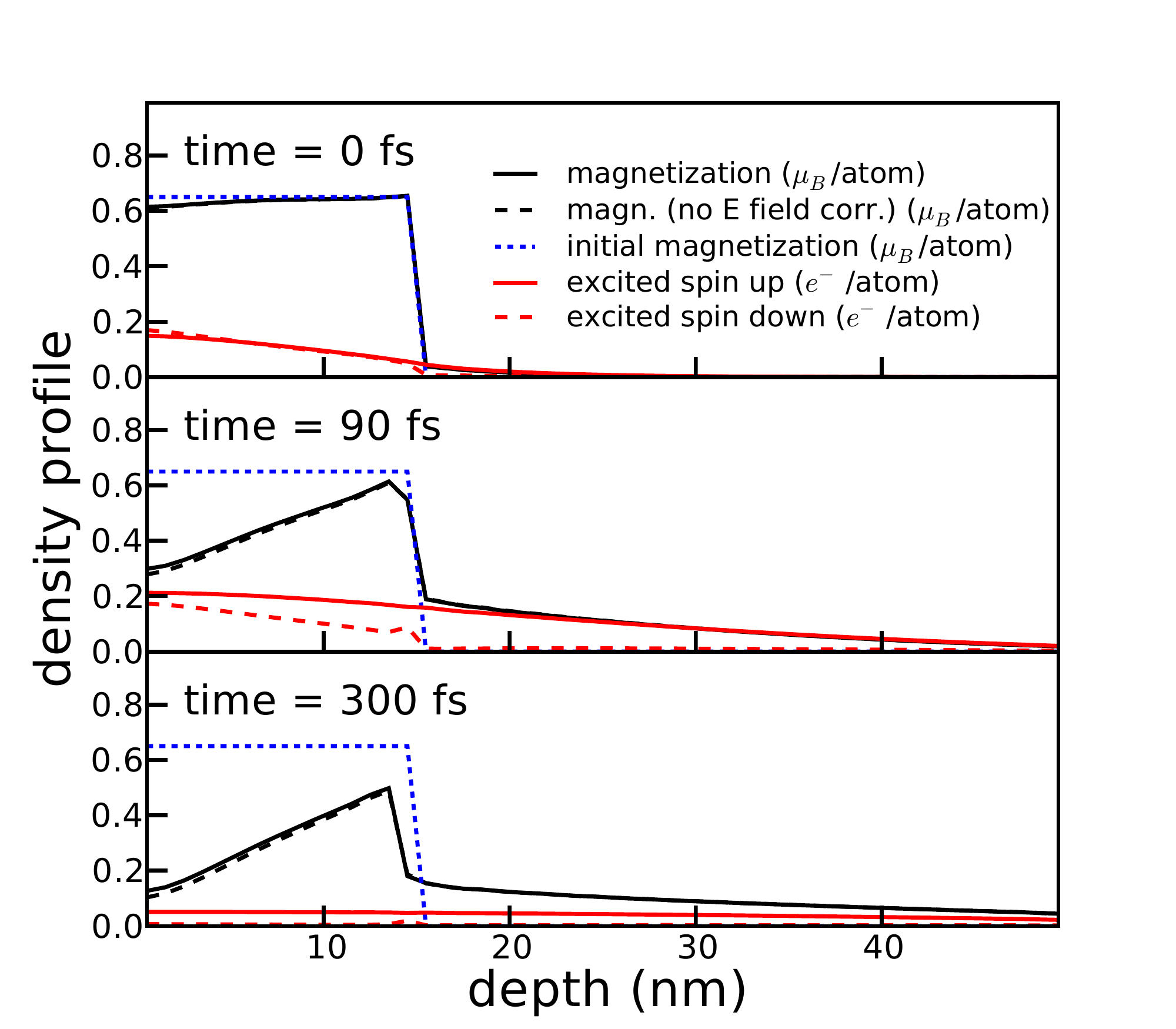}
 \caption{(Color online) Calculated spatial magnetization profile of Ni at three times caused by laser excitation (at $t$=0 fs). The resulting magnetization profile is given by the full curve, the initial one by the dotted curve. The electric field correction is illustrated by the dashed curve.
 The surface of the film with the vacuum is at 0 nm depth, the Ni film extends up to 15 nm depth,  the remaining is the Al film. 
 }
 \label{fig:magnprof}
\end{figure}

 To assess how much demagnetization such process can generate we solve Eq.\ (\ref{eq:continuity_eq_total_density}) numerically for the case of a 15-nm Ni film on an Al substrate. This system was recently investigated by Stamm {\it et al.} \cite{stamm07}. The energy dependence of the spin-polarized densities in Eq.\ (\ref{eq:continuity_eq_total_density}) is discretized in 12 channels from 0 up to 1.5 eV, the pump-laser energy. The interface with the vacuum is treated as a reflecting surface, but the excited carriers in the Ni film can penetrate the Al. A pump laser with a temporal profile of 60 fs (full-width at half-maximum) is adopted. Furthermore, excitation rates are taken from Ref.\ \cite{stamm07},  the ratio of excited spin-up to spin-down electrons from \cite{oppeneer04}, and the spin-dependent inelastic lifetimes and velocities from Ref.~\cite{zhukov06}. Unfortunately, Ref.~\cite{zhukov06} does not provide inelastic lifetimes for  excited electrons with low energies (i.e., from the Fermi level to 0.5 eV above). We made a linear extrapolation to estimate lifetimes for the lowest energies; the influence of this approximation is discussed further below, together with those of other approximations. The elastic lifetime contribution can be sample dependent. We used as elastic inverse lifetimes $100\%$ of the inelastic ones. Values from $20\%$ to $200\%$ have given no relevant changes in the demagnetization.

\begin{figure}
 \includegraphics[trim = 2mm 0mm 17mm 15mm, clip, width=0.45\textwidth]{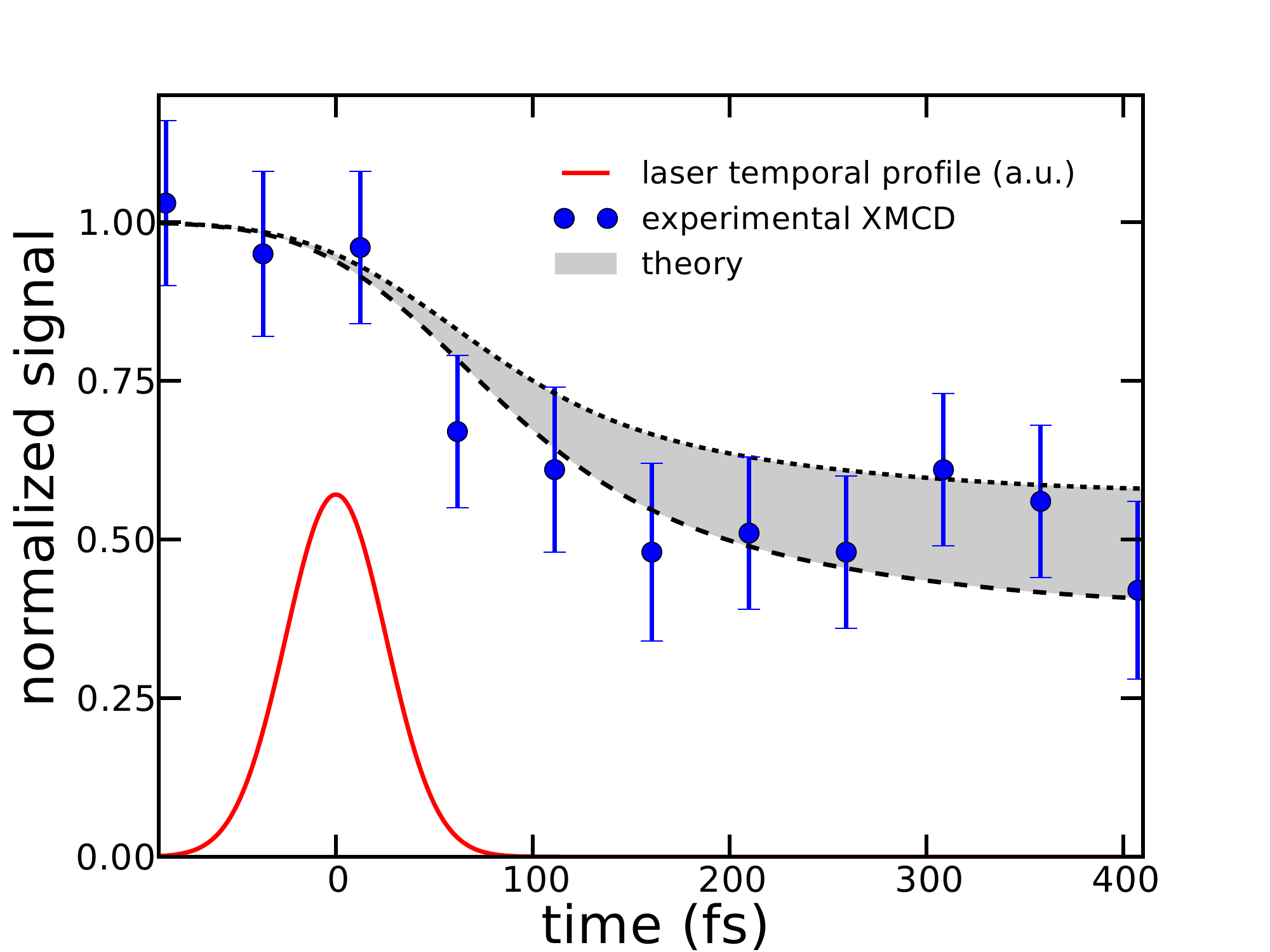}
 \caption{(Color online) Computed laser-induced demagnetization in Ni.
 The shaded area shows where the theoretical result is expected to be (depending on the inelastic lifetime). 
 For comparison we also show recent experimental XMCD data \cite{stamm07}. 
 The used time-structure of the laser pulse (in a.u.) is depicted by the red (online) line.}
\label{fig:timemagn}
\end{figure}

 Fig.\ \ref{fig:magnprof} presents computed spatial  ($z$ coordinate) profiles at three times. The maximum of the excitation laser is at 0 fs, but due to its temporal profile (cf.\ Fig.\ \ref{fig:timemagn}) there is already an effect visible at $t$=0 fs. The laser excitation is spin-conserving, hence in itself it does not cause any magnetization change. But due to spin-dependent fast transport of excited electrons the magnetization (in $\mu_B$/atom) begins to deviate from the initial one in the surface region. This process continues to $t$=90 fs, where the magnetization in the near-surface area has become considerably reduced; the super-diffusive flow of spin-up electrons into the Al film causes a magnetization of the Al film. At $t$=300 fs, well after the pump-laser has vanished, the continuing motion of excited electrons has created a further demagnetization, but the densities of excited carriers are already quite reduced. The latter stems from inelastic collisions which an excited electron suffers until it has lost its energy, i.e. the process of thermalization. Due to scattering in the Al layer, there is a back-flow from the Al that creates a minority spin accumulation on the Ni side, leading to a stronger demagnetization at the interface, visible for $t$=300 fs.

 The carrier flow creates charged regions that generate an electric field. This field is not negligible for laser fluences that give demagnetizations of the order of tens of percent. It will, however, act equally on both spin channels and on all conduction electrons, not only the excited ones that are a small percentage. As an extreme scenario we computed the back-flow necessary to compensate the free charge generated by the super-diffusive transport; the results are shown in Fig.\ \ref{fig:magnprof}. Its influence on the magnetization profile is indeed small.

 Next, we study the calculated time-evolution of the magnetization in Ni and compare to results obtained in a recent pump-probe experiment \cite{stamm07}, in which the x-ray magnetic circular dichroism (XMCD) was used to probe the magnetization. The time-evolution of the magnetization is shown in Fig.\ \ref{fig:timemagn}. The computed magnetization is convoluted with the probing beam's temporal profile (FWHM=100 fs) and averaged over the Ni-film thickness, because the XMCD signal is sensitive to the Ni only and almost uniformly probes the whole film thickness. The predicted demagnetization is in reasonable agreement with the experimental XMCD data. After 200 fs the magnetization is reduced to about 50\%. The XMCD experiment seems to indicate a somewhat faster demagnetization, but we must be careful when directly comparing the absolute values of measured and calculated magnetization. It has been shown that, while demagnetization is occurring, the XMCD signal also contains artifacts due to state-blocking effects \cite{carva09}. The shaded area gives the temporal behavior computed for two different zero-energy extrapolations of the lifetime data \cite{zhukov06}. A linear extrapolation of lifetimes for energies near the Fermi level gives an inelastic majority spin lifetime of 140 fs; a rigid cut-off of the low-energy lifetime gives a value of 50 fs. The former corresponds to a stronger, the latter to a weaker demagnetization ($\sim$40\%). Using longer inelastic lifetimes does not lead to a notable increase of the demagnetization.

 We note that our model does not include two effects that might give additional contributions: the generation of Auger electrons and the response of the material to the new electronic state. The first effect provides a delayed source of fast electrons, improving the effectiveness of the transport-driven demagnetization. The second effect is more complicated. The inelastic lifetimes might be altered by the new electronic distribution which may reduce the effectiveness of the process. 
	
 A confirmation of our theory is that it explains the striking difference between magnetic metals and magnetic dielectrics. The latter exhibit much slower demagnetization times \cite{muller09,kirilyuk10}. As the super-diffusive transport processes are inhibited in insulators, the demagnetization has to evolve through the much slower dynamics of spin-flip processes.
	
 In contrast to Ni, a very slow laser-induced demagnetization has been observed for metallic $4f$ ferromagnets as Gd \cite{melnikov08}. Gd is different from metallic $3d$ ferromagnets because the spin-polarized $4f$ states--which contribute 7 $\mu_B$ to the Gd moment--are well localized and not reached by the pump laser. Only the Gd $5d$ band electrons near the Fermi energy are excited by the pump laser, but these electrons contribute only 0.55 $\mu_B$ to the atomic moment. Our theory is consistent with the two-time-scales demagnetization in Gd, because the super-diffusive transport of laser-excited $5d$ electrons would create a small, fast $5d$ demagnetization, but then the laser-imparted energy needs to be transferred from the $d$ electrons to the lattice and the $f$ electrons to achieve sizable demagnetization. Especially the latter process is slow ($\sim$80 ps, \cite{melnikov08}), as it involves spin-flip phonon scattering.

 Malinowski {\it et al.} \cite{malinowski08} observed a demagnetization contribution due to spin transport, however, they did not have a detailed description of the process to quantify its amount. We have developed here a model that can explain laser-driven fs demagnetization on the basis of spin angular-momentum conserving super-diffusive transport. Numerical solutions of the transport equations show that a substantial demagnetization of $\sim$50\% is created within 200 fs, without the need to invoke any spin-flip channels. At this stage we cannot exclude other demagnetization mechanisms, but our calculations show that super-diffusive processes play a main role, and can explain wholly the ultrafast demagnetization process during the first few hundred femtoseconds. On longer timescales other well-known effects that are not included in this treatment but that can be modeled by Landau-Lifshitz-Gilbert dynamics \cite{Kazantseva08} come into play. A combination of these approaches is required to treat the full fs to ps time-domain. A test of our theory could be to observe the flow of spin-polarized electrons in the non-magnetic substrate, which, for sufficiently thin substrates could be detectable by magneto-optical techniques.


\acknowledgments
{We thank H. D{\"u}rr, I. Di Marco, and J. Rusz for valuable discussions. 
This work has been supported by the Swedish Research Council (VR), by FP7
EU-ITN ``FANTOMAS",  the G. Gustafsson Foundation,
and the Swedish National  Infrastructure for Computing (SNIC).
}


\end{document}